\documentclass[12pt,a4paper,english]{article}
\usepackage[T1]{fontenc}
\usepackage[latin2]{inputenc}
\usepackage[english]{babel}
\usepackage{amsmath,amssymb,pstricks}
\usepackage{amsfonts}
\usepackage{indentfirst}
\usepackage[dvips]{graphicx}

\begin{document}

\sloppy


\begin{center}
\begin{Large} 
{ \bf Dodging the crisis of folding proteins with knots}
\end{Large}

\bigskip

\bigskip

\bigskip

Joanna I. Su{\l}kowska$^{1,2}$, Piotr Su{\l}kowski$^{\,3,4}$, Jos{\'e} N. Onuchic$^1$

\bigskip

\bigskip

\emph{$^1$Center for Theoretical Biological Physics, University of California San Diego, \\
Gilman Drive 9500, La Jolla 92037}

\medskip

\emph{$^2$Institute of Physics, Polish Academy of Sciences, \\ Al. Lotnik\'ow 32/46, 02-668 
Warsaw, Poland}

\medskip

\emph{$^3$Physikalisches Institut der Universit{\"a}t Bonn and Bethe Center for Theoretical 
Physics, 
\\ Nussallee 12, 53115 Bonn, Germany}

\medskip

\emph{$^4$So{\l}tan Institute for Nuclear Studies, Ho\.za 69, 00-681 Warsaw, Poland}

\bigskip

\bigskip

\smallskip
 \vskip .6in \centerline{\bf Abstract}
\smallskip

\end{center}

Proteins with nontrivial topology, containing knots and slipknots, have
the ability to fold to their native states without any additional  
external forces invoked.
A mechanism is suggested for folding of these proteins, such as YibK  
and YbeA, which involves
an intermediate configuration with a slipknot. It elucidates the role  
of topological barriers and backtracking
during the folding event. It also illustrates that native contacts are  
sufficient to guarantee folding in around 1-2\% of the simulations,
and how slipknot intermediates
are needed to reduce the topological bottlenecks. As expected,  
simulations of proteins with similar structure but with knot 
removed  fold much more efficiently, clearly demonstrating the origin of these  
topological barriers.
Although these studies are based on a simple coarse-grained model,  
they are already able
to extract some of the underlying principles governing folding in such complex  
topologies.

\bigskip

\textsf{knots | slipknots | proteins | folding | backtracking | molecular dynamics }

\bigskip

\newpage

During the last two decades, a joint theoretical and experimental
effort has largely advanced 
the quantitative understanding of the protein folding
mechanism. Most small and intermediate size proteins  live on a minimally
frustrated funnel-like energy landscape, which allows fast and robust
folding \cite{Onuchic_1995,Onuchic_2000,Onuchic_2004}. 
Since proteins have been able to solve the energy problem, the final
challenge is the structural complexity of the protein folding motifs. 
Most proteins avoid complex topologies, but recent discoveries have shown that
some proteins are actually able to fold into non-trivial topologies where the
main chain folds into a knotted conformation \cite{Taylor_2000,Virnau,Taylor_2007}. 
Although these ``knotted'' folding motifs have been observed, we still have to face 
the challenging question of how the protein overcomes the 
kinetic barrier associated with the search of the knotted conformation. We
suggest a possible mechanism where the knot formation is preceded by a
conformation called a ``slipknot''. 
A slipknot is topologically similar to a knot, expect that an internal knot is effectively
undone as the pathway of the backbone folds back upon itself.  
The fact that such slipknots have already been observed in some protein 
final structures \cite{Yeates_2007} adds support to this suggestion.

\begin{figure}[htb]
\begin{center}
\includegraphics[width=0.55\textwidth]{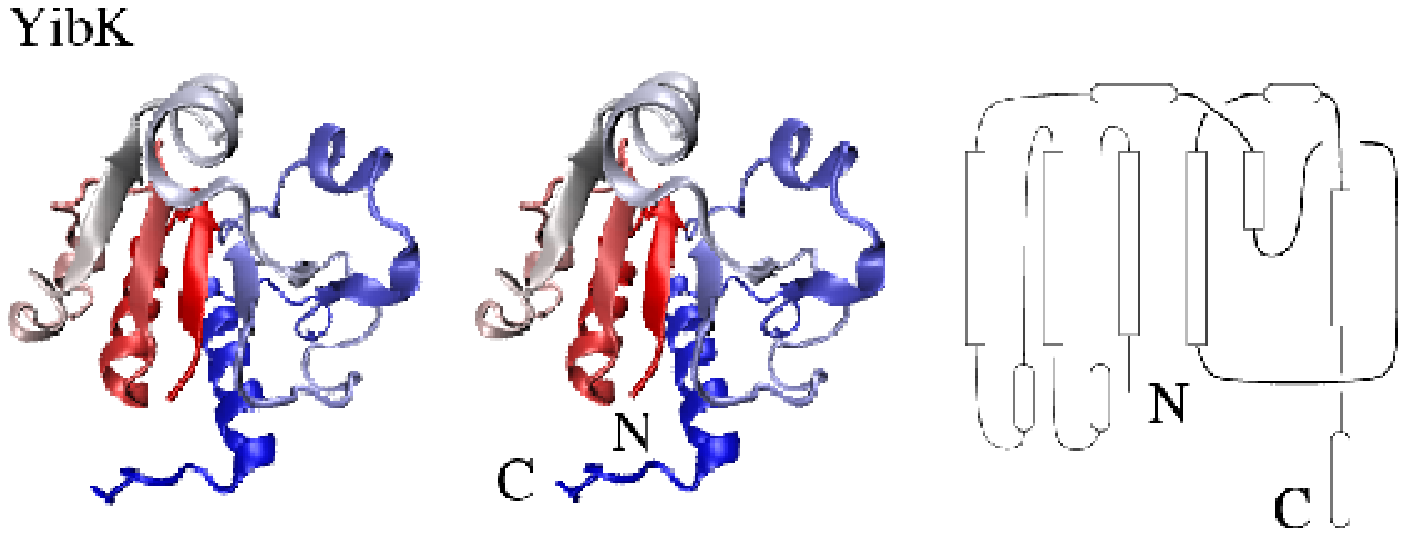},
\includegraphics[width=0.55\textwidth]{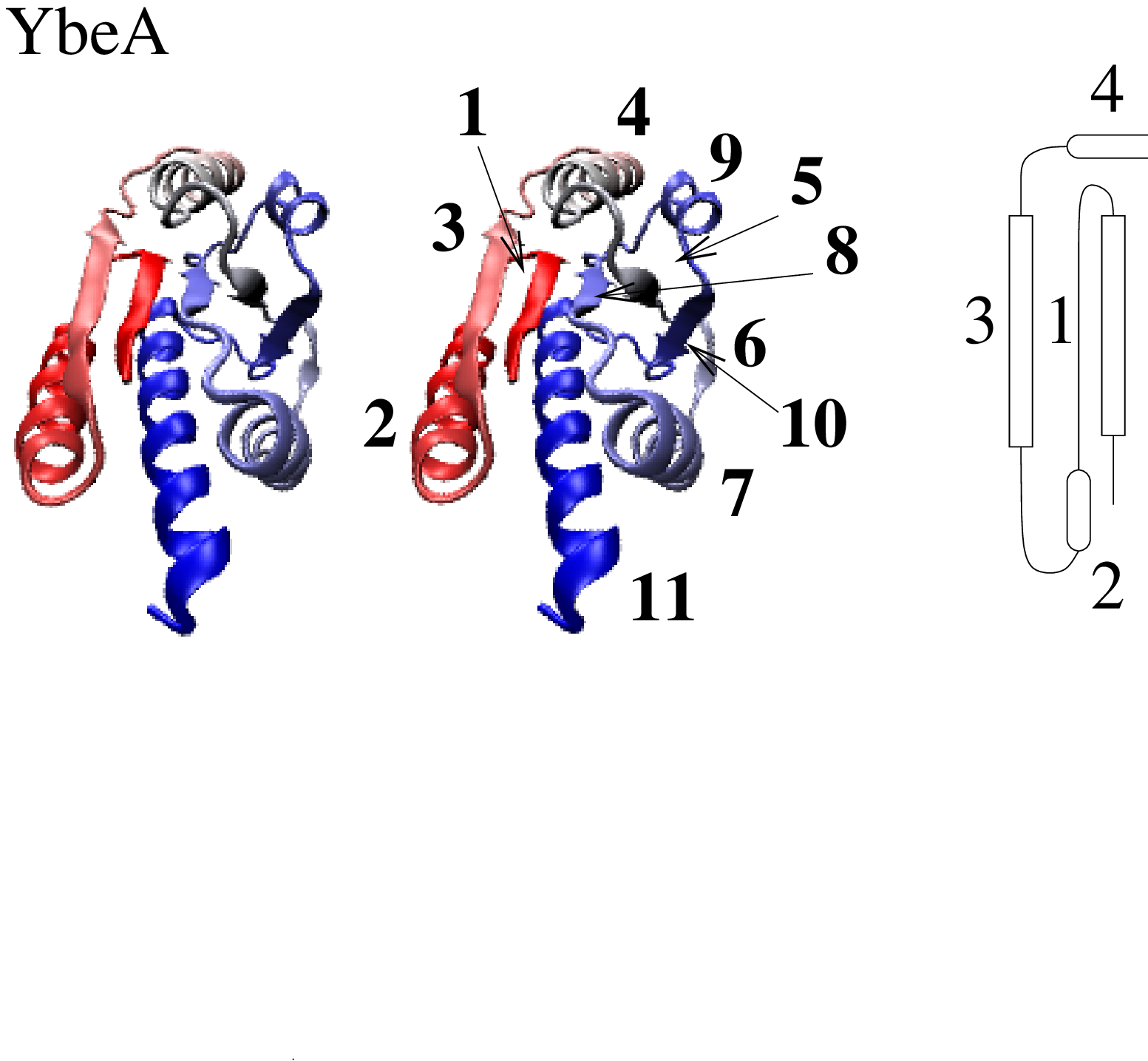}
\caption{Structure of knotted proteins.
Top panel -- stereo view of structure of knotted protein  YibK (1j85) (left)
and its schematic structure (right).
Bottom panel -- stereoview of YbeA (1od6) and its schematic structure (left).
Details of proteins structure are described in \emph{Appendix}.
} \label{zwijanie-rys}
\end{center}
\end{figure}

This folding mechanism is explored in the context of the two 
most experimentally investigated knotted families of proteins, 
Haemophilus influenzae YibK and Escherichia coli YbeA,  which are
homodimeric $\alpha/\beta$-knot methylotransferases (MTases). 
A schematic representation of these proteins is shown in Fig. \ref{zwijanie-rys}. 
It has been shown experimentally that both these proteins unfold spontaneously and reversibly upon
addition of chemical denaturant \cite{mallam_2005,mallam_2006,mallam_2007a,mallam_2007b} 
and they are able to fold even when additional domains are attached to one 
or both termini \cite{mallam_2008}. In very recent experimental work \cite{mallam_2008a}, based on analysis of 
the effect of mutations in the knotted region of the protein, a folding model for 
YibK was also proposed. In this model the threading of the polypeptide 
chain and formation of the native structure in the 
knotted region can occur independently as successive events. 
These results alone, though, are not sufficient to explain the folding mechanism.
To complement the experimental information, we have devised a theoretical computational strategy.
Simulations are performed for three proteins using structure-based coarse-grained models,
based on the $\alpha$/$\beta$-knot superfamily YibK and YbeA.

In the present study we consider knots of simplest type, referred to as a trefoil or $3_1$ knot.
It that consists of a loop through which one end of a chain is threaded. 
In principle there might be three possible mechanisms leading to the creation of such a knot. 
The most straightforward one would require just two steps: creating a loop and threading one end 
through it. The second mechanism is more complicated and involves an intermediate step with a slipknot. 
The third and final possibility would involve the creation of an ensemble of loose random knots in the
first stage, which may turn into deeper knots after a relatively longer time. 
Analysis of simulated folding trajectories provides the necessary insight on this complex folding event.
The results indicate that the  folding of YibK and YbeA proceeds according to
the second mechanism, through a slipknot
intermediate configuration. This is consistent with earlier theoretical 
observations \cite{Taylor_2007,Yeates_2007}. 
Fig. \ref{1o6d_path} provides a detailed description of the suggested folding mechanism.

Understanding the folding mechanism for these knotted proteins provides  
the tools to explore additional complex folds. For example, one can extend 
these studies to proteins that do not fully knot but form a slipknot in the native state. 
Some slipknotted proteins have a simple folding motif such as crenarchaeal viruses AFV3-109 
\cite{Keller} but others, such as thymidine kinase \cite{Gardberg}, 
are longer and have a more complex folding mechanism.

\section{Results - Folding Knotted Proteins}

\begin{figure}[htb]
\begin{center}
\includegraphics[width=0.75\textwidth]{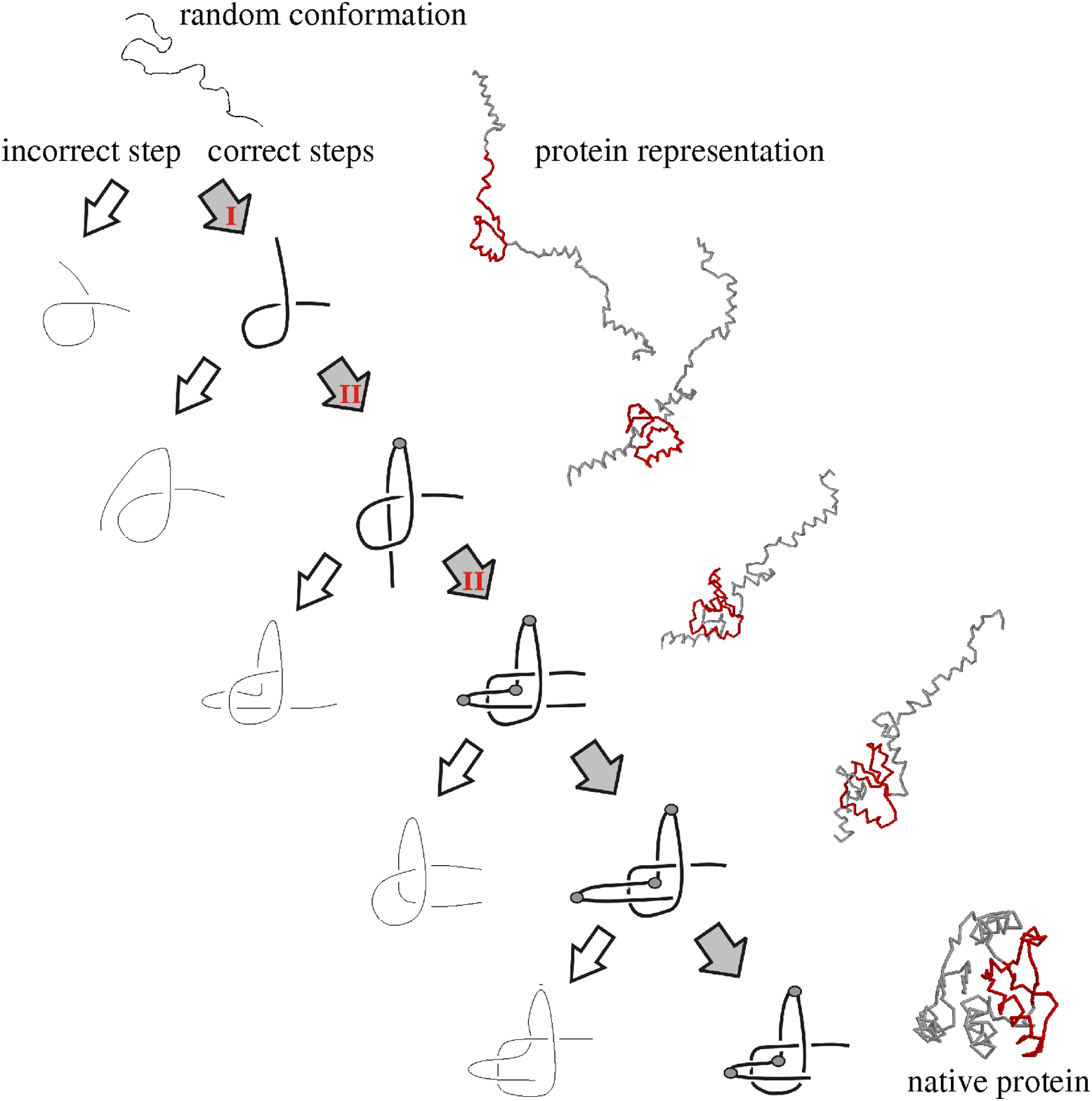}
\caption{The folding route 
that leads to the native knotted configuration $3_1$ through an intermediate configuration with a slipknot. 
The pathway is represented by a series of configurations, 
starting in the top left (a random conformation) and proceeding towards the right bottom.
The path involves five essential steps, the first three of them correspond to the Reidemeister moves I and II
(denoted on appropriate arrows). A typical representation of the protein
conformations are shown to the right of the associated mechanistic step. Native-like locations
of the knot are shown in red.  Each step is characterized by the
appearance of new structural elements such as loops and hooks. During the
folding route incorrect steps may occur. Examples of some  incorrect
configurations, which are kinetic traps, are represented by thin lines
(sketched on the left). These traps act as topological barriers \cite{norc} 
and escaping from them requires a backtracking mechanism similar to 
the ones we have observed in regular folding \cite{frust}.
Grey dots represent amino acid 102 in the third step plus amino acids 
114 and 124 in the following steps. Special emphasis is given to them 
since they are associated to sharp turns,  which are required to achieve the knotted state.
This folding process is shown in the attached movie, 
whose details are described in \emph{Appendix}.
} \label{1o6d_path}
\end{center}
\end{figure}

The folding pathways which lead to the
knotted conformation are observed in our simulations through an intermediate configuration containing a
slipknot. This route is shown schematically in Fig. \ref{1o6d_path} 
and discussed below. Analysis of this suggested folding mechanism requires that we divide this route 
into a few geometrically distinctive  steps. A step may consist, for example, of threading the chain 
through a loop created  in a separate region of the sequence.
This representation allows us to describe  long folding
trajectories, consisting of thousands steps in our simulations, by a sequence of the essential 
intermediate ensemble of 
configurations. When projected on a plane, certain transitions between such two adjacent configurations 
can be identified with so called Reidemeister moves. The Reidemeister moves change a relative location of some
strands (when projected on a plane), but do not change the type of the knot. There are three such moves, denoted I, II and III, that
involve respectively one, two and three strands. They are described in detail in \emph{Appendix}.

\subsection{Folding proteins in the $\alpha/\beta$-knot superfamily into a trefoil knot}

Our analysis is based on the simulations of the  $\alpha/\beta$-knotted proteins YibK (PDB code 1j85) and YbeA (PDB code 1o6d,
1vho), shown in Fig. \ref{zwijanie-rys}, which are homodimers in their native configuration.
According to the experimental results \cite{mallam_2006,mallam_2007a}
they fold via a few intermediate steps reaching the stable monomer states with considerable structure.
Only the last step involves formation of a dimer. This allows us to study these proteins as monomers.
Notice that YibK shares only 19\% sequence identity with YbeA \cite{mallam_2007a}.
Folding simulations were performed at temperatures slightly above the folding temperature $T_{f}$,
starting from random conformations.
Simulations were run utilizing a standard $C_{\alpha}$ structure-based model (sometimes called Go Model)
\cite{Clementi,Cieplak_2003}.
For each protein (1j85, 1o6d, and 1vho) we observed at least 10 successful folding trajectories; 
around 1-2\% of all routes succeeded to reach the native knotted states.
These successful trajectories choose between two parallel folding trajectories similarly 
to what has been proposed experimentally \cite{mallam_2006,mallam_2007a}. 
Both trajectories are characterized by an intermediate slipknot configuration but in one case 
it is formed early in the folding process while, in the other one, it is a late event. 
Fig. S2 in \emph{Appendix} shows the fraction of native contacts $Q$ at the moment when
the slipknot is being created, i.e. when its hook-region starts to be threaded through the loop 66-96.
In the most common trajectory the structure elements located closer to the N terminal
of the protein fold in the final stages of the folding process around $Q \simeq 0.8$.
In the alternative trajectory, the N terminal folds in the early stages of folding, thus when the slipknot 
is being created $Q \simeq 0.6$. In fact there is also an additional but very
rare possibility (which we observed only once) 
of formation a knot by threading the terminus C through the loop 66-96 without a slipknot intermediate.

For clarity we now discuss in detail one folding trajectory for 1o6d, shown schematically in Fig. \ref{1o6d_path}
(this is the smallest of all proteins being analyzed and therefore had the largest number of successful runs). 
Although we focus on this single example, this trajectory is similar to other proteins with a knot $3_1$, as 1vho, 1j85.
This folding route consists of six distinctive intermediate configurations. During the first transition a loop is created 
(the Reidemeister move I). This loop extends between amino acids 66 and 96 and remains formed during the entire  folding
event. Following this step, a further region of the chain gets close to the loop, 
requiring a hook formation (the Reidemeister move II). This hook is then threaded through the loop 
(also move II), creating the  slipknot in the fourth intermediate.
The slipknot is then transformed into a knot by pulling one termini through the loop,
which may be viewed as a two-step process, with an intermediate 5 and final configuration 6. 
There are no Reidemeister moves corresponding to these last two steps, because they could not happen for 
a closed chain without cutting it.

As argued above, the creation of a knot requires several turns to occur at the right place and at the right time order. 
In the trajectory  shown in Fig. \ref{1o6d_path} these turns are represented by grey dots. 
The first one occurs at  amino acid 102, which redirects the protein 
chain towards the loop 66-96. Then two other turns at positions 114 and 125 are needed for the hook formation. 
This hook is threaded through the loop 66-96. Finally the termini 147 is pulled through
this loop which results in the $3_1$ knot.

To understand the knot formation, we investigate the order that native contacts are formed and, sometimes, broken and reformed (backtracking) during the folding event. Correlation between formation of some native contacts with the backtracking of other ones can teach us about topological bottlenecks.
This information can be extracted from Fig. \ref{fig-contacts}, where the time dependence of the number of formed native contacts between relevant structural elements of the protein knot is shown. Initially we focus on the formation of the loop between residues 66 and 96  that requires the creation of contacts between $\beta$-strands 5 and 8. This can only be achieved by the pre-formation of the contacts between 5 and 10, which is followed by a simultaneous destruction of these contacts and formation of the ones between 5 and 8. During this loop formation, contacts between strands 6 and 10 are also formed, which are needed to create the slipknot conformation.
The formation and destruction (around 180 timesteps) of contacts between strands 5 and 10  is an example of backtracking. Backtracking also plays a role in the formation of contacts between 5 and 9 and between 6 and 9. The order of contact formation is crucial, otherwise knotting is not possible.

Backtracking is also a mechanism to escape from kinetic traps as described in Fig. \ref{1o6d_path}.
For example, if  8-11 contacts are formed early, the creation of a loop between residues 66 and 96 becomes impossible. The 8-11 contacts 
need to be broken before this loop can be  created. A similar situation occurs at 
the final stage of knot creation that  involves threading the helix 11 through the loop. During this process, contacts between helices 7 and 11 may be accidentally created. They need to be  broken until further threading can proceed. 

\subsection{Energetic heterogeneity enhancing folding ability} 

The results above demonstrate that making only the native contacts attractive is sufficient to fold the protein 
into its native conformation.
The percentage of successful trajectories, however, is then very small -- only about 0.1\%. Analysis of this data suggests that this success rate can be increased by improving the time-order of contacts formation and therefore reducing backtracking. We utilized different strength for a few selected native interactions to achieve this folding improvement. Backtracking gets reduced but the overall folding mechanism remains the same as described in Fig. \ref{1o6d_path}.

\begin{figure}[htb]
\begin{center}
\includegraphics[width=0.75\textwidth]{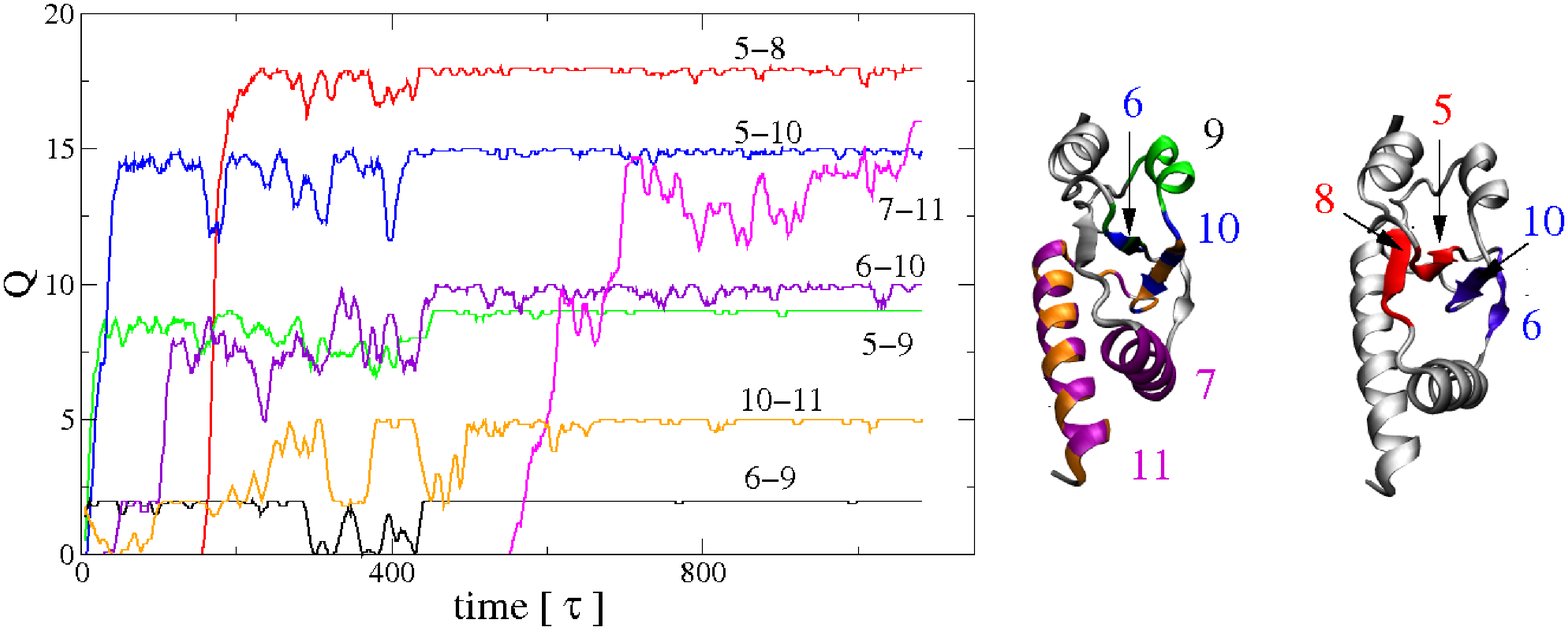}
\caption{Left: the time dependence of the number of formed native contacts between relevant structural elements of the protein knot, 1od6.
Right: cartons representation of knotted part of the protein 1od6. Interacting regions of the protein 1o6d are colored accordingly 
to the ones used in each trajectory.
For clarity, if there are native interactions of one secondary structure
element with two different parts of the protein, they are shown on
two separate structures.
Backtracking is observed at different levels for all trajectories.  
The contacts between $\beta$ strands 5-8 are colored red, the ones between $\beta$ strands 6-10 are magenta, and the ones for the turn 97-105 are orange.
Kinetic studies were performed using overdamped Langevin dynamics. Timesteps have been chosen to have typical folding times of about 1000 steps.}
\label{fig-contacts}
\end{center}
\end{figure}

Guided by the results of Fig. \ref{fig-contacts} which identifies the structural regions responsible for enhancing folding, we modified some of our contact interactions.  We increased the strength of contacts  between $\beta$ strands 5 and 8. This modification facilitates the formation of the loop between residues 66 and 96 and also the attachment of the  C terminal to this loop, which leads to the slipknot
conformation involving the contacts between strands 6 and 10. We also slightly reduced the interaction  energy between
the C-terminal and the loop. Rapid formation of those contacts results in problematically fast
blocking of the necessary  backtracking and threading of the C-terminal across the loop. Simulations with this ``optimal'' model succeeded in folding into the knotted structure 29 times in 2500 trials. Further increasing the contact strength in these regions does not keep improving knot formation. Since backtracking is also needed during the knotting process, if those increases are too large, they will lead to more kinetic traps.
For further analysis on how this "optimal" set of contacts have been determined see the  \emph{Appendix}.

\subsection{Extension of the protein chain and a rebuilt protein without knot structure}

We also analyzed how folding properties of these proteins were affected when additional chains (tails)
were attached to one or both termini following the idea of experiments reported in \cite{mallam_2008}.
Due to computational limitations we restricted our studies  to purely flexible homopolymer tails  built of no more than 12 residues.

The addition of these tails not only did not restrict,  but sometimes even increased the number of
correctly formed knotted native states. The last step in the folding process, which involves pulling the entire tail
through the loop, is slower than in the regular protein.  Nonetheless, the overall folding time is similar in both cases.

The final test which can clarify the folding ability of the knotted protein (1j85) 
and the presence of a topological barrier is 
to use a modified protein in which the knot is absent. To engineer such a protein
with a trivial topology, it is sufficient to change the crossing of 
protein chain between 78-85 and 120-125 amino acids using methods from \cite{Gront_2007,Gront_2008}.
The folding ability of such an untighted structure increases to a 73\% success for equally long folding runs.
This means that a simple change of topology 
eliminates an otherwise very high topological barrier.



\section{Results - Folding Proteins with Slipknots}

\begin{figure}[htb]
\begin{center}
\includegraphics[width=0.75\textwidth]{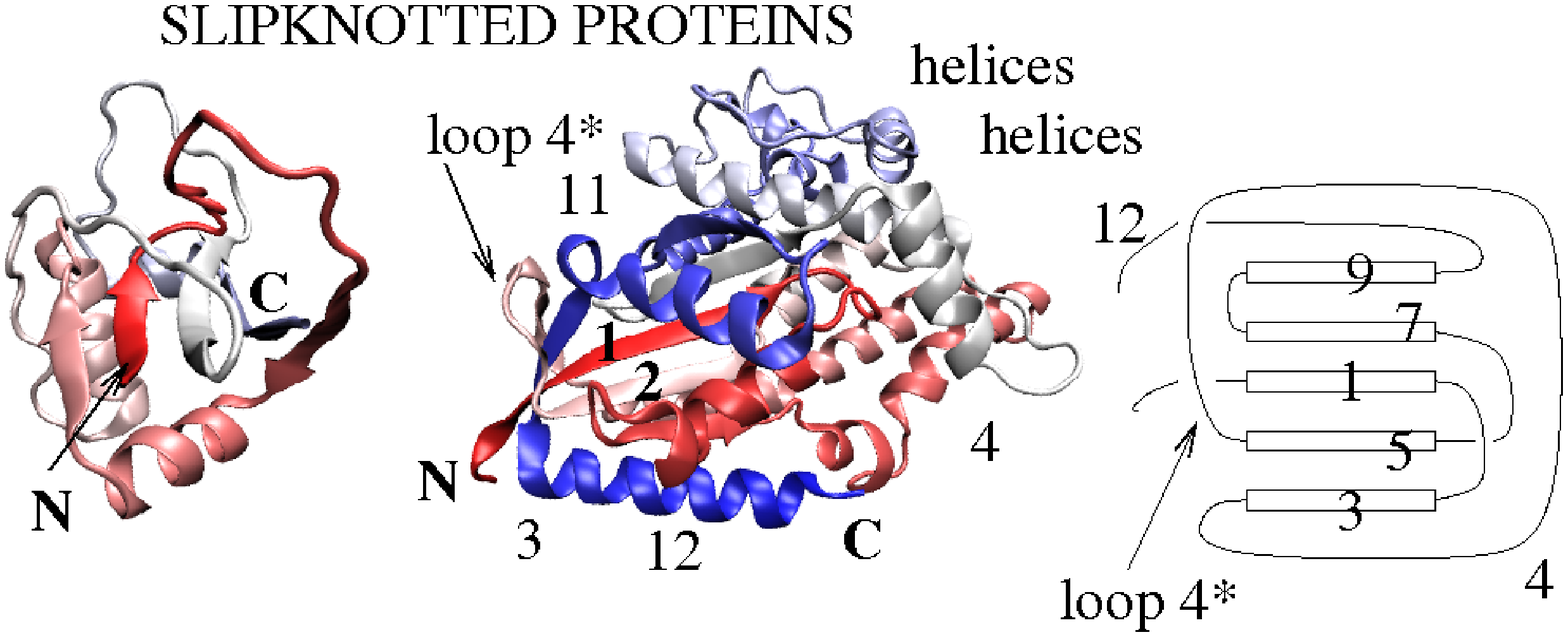}
\caption{
Structure of two proteins with slipknots:
crenarchaeal viruses AFV3-109 (2j6b) (left) and  a thymidine kinase (1p6x) (middle).
Right: a schematic representation of the structure of 1p6x.
Stereo-views of proteins is shown in the Appendix.}
\label{zwijanie-rys-slip}
\end{center}
\end{figure}

So far we have discussed  proteins that have a $3_1$
knot in the native conformation.  Our suggested folding mechanism  via a slipknot intermediate configuration gets further support from the fact that in a different class, proteins actually contain slipknots in their native states 
\cite{Taylor_2007,Yeates_2007}. 
Comparing their folding mechanism to our initial results provides further understanding of our proposed model.
We focus on a relatively short protein 2j6b (consisting of 109 amino acids)
and a much longer one 1p6x (333 amino acids), which are shown in Fig. \ref{zwijanie-rys-slip}.
Again, a model that includes only attractive interactions for native contacts is sufficient to fold these proteins.
Although both these proteins contain slipknots, their folding trajectories
are different. Folding of 2j6b is composed simply by the first three steps in Fig. \ref{1o6d_path}.
This independently provides strong support for our proposed folding mechanism for the knotted proteins.
Formation of a slipknot in 1p6x is more complicated and it is described below.

\begin{figure}[htb]
\begin{center}
\includegraphics[width=0.6\textwidth]{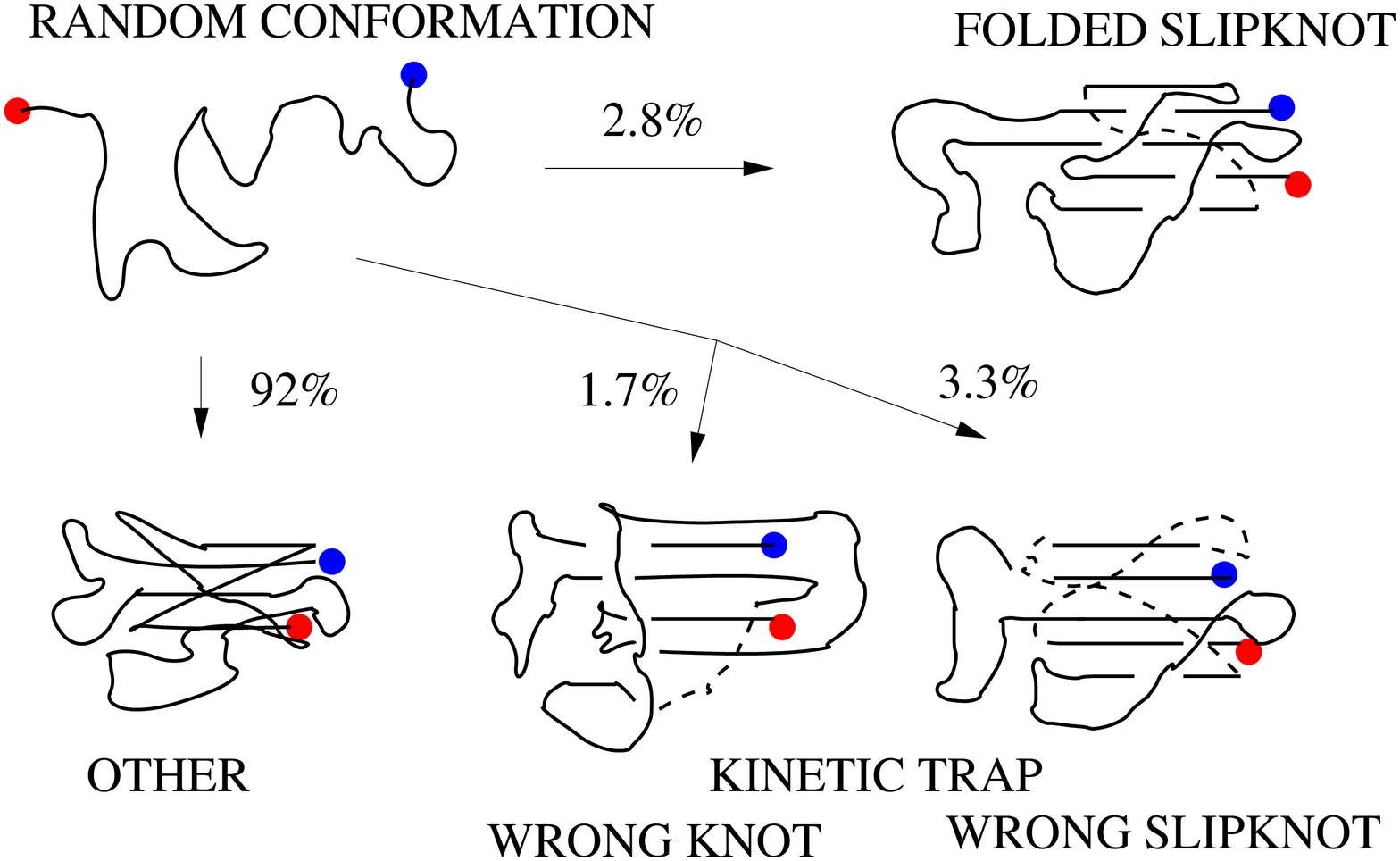}
\caption{Schematic representation of the four different ensembles of final conformations observed in  1000 folding runs of 
the slipknotted protein 2j6b. Simulations were performed at temperatures slightly  above $T_{f}$. 
Only the structure in the top right corresponds to the correctly folded protein with the correct slipknot and all native contacts.} 
\label{zwijanie-rys-2j6b}
\end{center}
\end{figure}

The folding mechanism for the short protein 2j6b involves creation of a loop
through which a hook is subsequently threaded. As discussed above, its folding is analogous to the first three steps
of folding of 1o6d.  Again, folding simulations were run at temperatures slightly above $T_{f}$.
Based on 1000 trajectories, we found four different ensembles of final conformations,  
shown in  Fig. \ref{zwijanie-rys-2j6b}.
Only 2.8\% of these trajectories reached the correct folding basin using three slightly different routes.
A few conformations ($\sim 0.5 \%$) in the "OTHER" basin in  Fig. \ref{zwijanie-rys-2j6b}
 have most of the native contacts formed but do not form  the slipknot.
Interestingly, the typical folding time for these trajectories is much longer than 
for such ones that reach the correct native folding.  This indicates that slipknot creation
should be a fast process, as long as the correct trajectory is followed.
We have also attached a tail to 2j6b to the termini which is closer to the
slipknot conformation. Similarly to the  knotted proteins, it did not substantially affect its folding properties. 
These extended proteins were able to fold into slipknot configurations in a similar fashion to the original protein.

We now discuss thymidine kinase (PDB code 1p6x), whose structure is described 
in detail in the \emph{Materials and Methods} section. 
We checked its folding ability at several temperatures above $T_f$, and for each one we ran 500 trajectories.
The highest probability of folding was observed at temperatures 18\% above $T_f$.   
In a set of 8000 trajectories, the correctly folded conformation including the slipknot
and all native contacts was reached only 11 times. In an additional 12 cases,  the slipknot 
was created but in a structure with an incorrectly folded bundle of 5 helices;
in these cases 92\% of native contacts where established.
In yet another set of 13 trajectories, although most of the native contacts were formed,  
a knot instead of a slipknot was observed in the final structure.
Knotted structures were reached in 5-20\% of the runs at each temperature and they had typically less 
than 73\% of native contacts.
All other final structures were some trivial kinetic traps or misfolded conformations.

Folding of the 1p6x protein, due to its large size and complexity, 
is more interesting than folding of the previously discussed proteins.
It can start from several nucleation sites. One of such sites includes a loop similar to the one reached after the first step in Fig.~2.
Notice that in the folding motif both termini are close to the loop. Also, during folding, both of them have to cross the loop.
This implies that, even though in the folding of 1p6x the steps from Fig. 2 are used, it proceeds in a different way than 
creation of a slipknot in 1o6d (where only one terminus had to be threaded through a similar loop). 
Therefore the folding mechanism is similar to the one shown in Fig.~2 but then the second terminus also has 
to thread through the loop.    

For the 11 correctly folded trajectories, we identified four possible folding routes \cite{Cieplak_2003}. 
They were classified  according to the instant during folding in which the slipknot contacts 
are formed  for the first time (see Fig. \ref{pathway-all}). 
Three of these four possible routes, albeit slightly different, 
have basically the folding mechanism described in Fig. 2 plus an additional step. 
As before their folding mechanism leads initially to the knot formation close to the N terminal. 
This is followed by threading the C terminal through the loop.
This final step gives rise to the slipknot conformation.
These three folding routes share several common features.
First, their nucleation sites, although they always include the loop, 
are composed by different additional regions of the protein \cite{norc}): 
bundles of alpha helices, a vicinity of the $\alpha$ helix at
the N terminal, or a huge loop close to the C terminal. 
Second, both termini enter the loop in configuration of a loosely packed hairpin.
Third, for all folding trajectories, backtracking is related to two topological 
barriers associated with the formation of the knot closest to the N terminal and to the  
 slipknot formation during last stages of folding.  

\begin{figure}[htb]
\begin{center}
\includegraphics[angle=-90,width=0.6\textwidth]{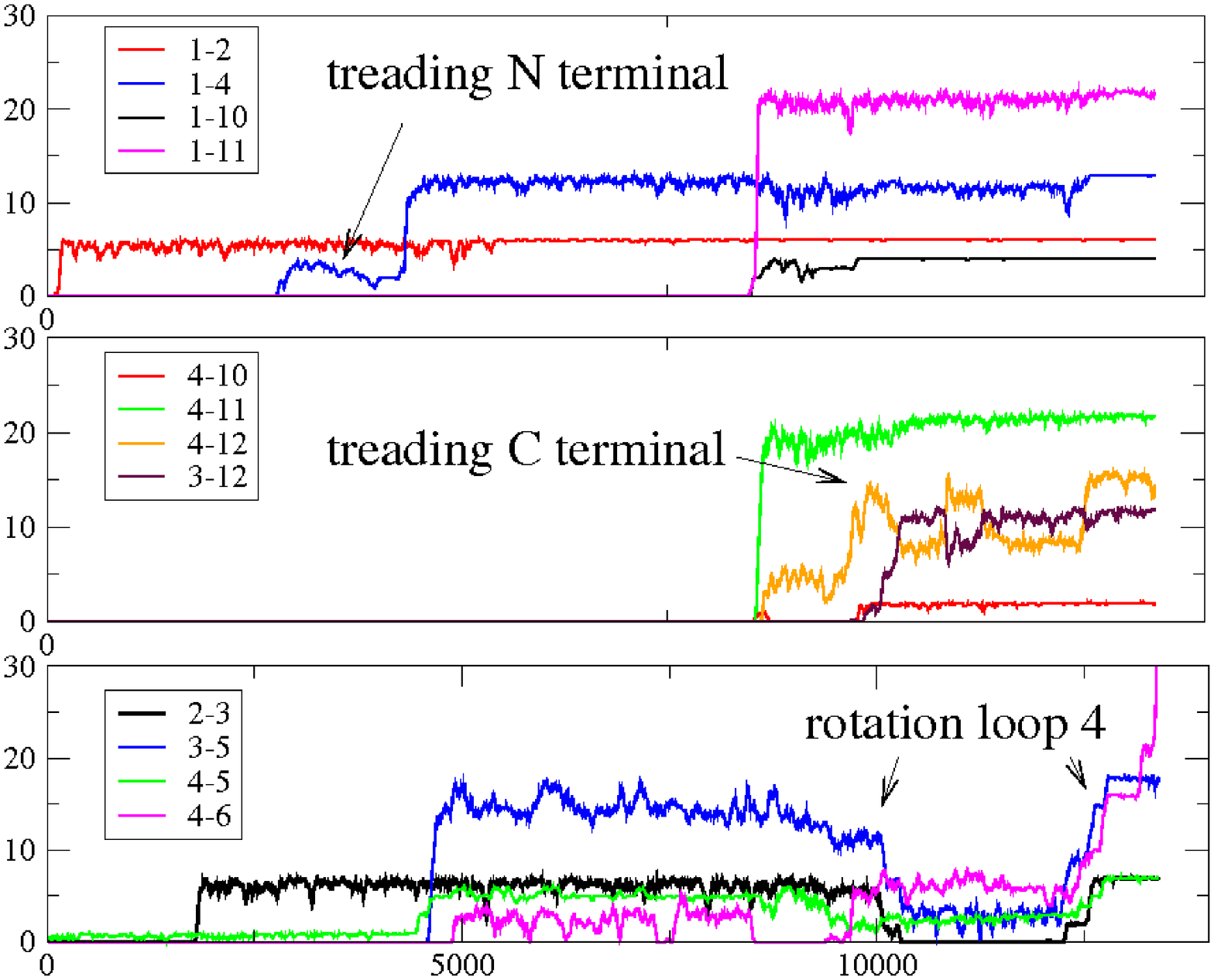}
\includegraphics[width=0.6\textwidth]{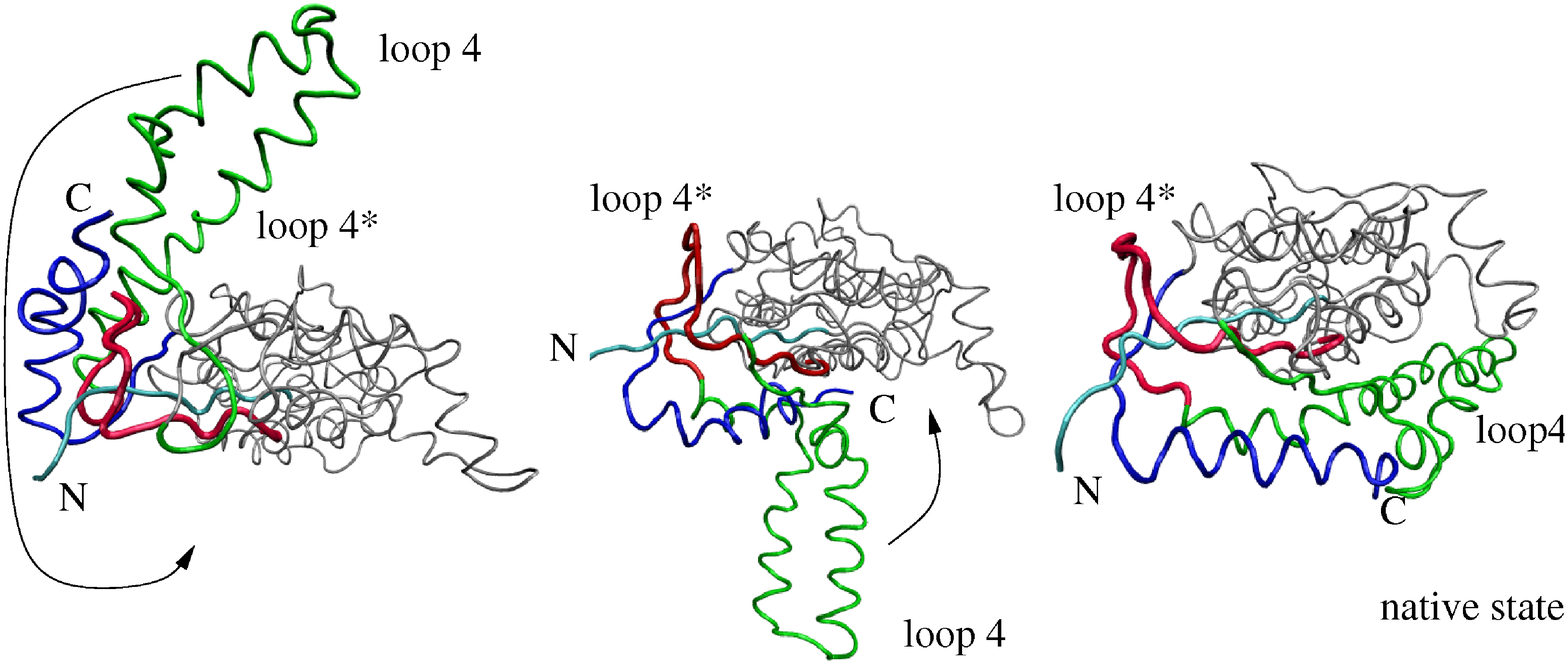}
\caption{The folding route for 1p6x (described as the fourth one in the text)
  which involves a rotation of the loop 4.
The three figures from the top, respectively: average number of contacts
in indicated sets of secondary structure around
N terminal, C terminal and loop 4 during the folding of 1p6x.
Bottom panel: steps which leads to a formation of a knot and slipknot structure
by the rotation of the loop 4. The steps are connected to the top panels in the following way.
In the first panel at time 2000 $\tau$ the knot is already partially formed, but not yet a slipknot, 
thus the loop 4 is still above the sheet 1-3-5. 
Then the backtracking follows, which involves $\beta$ strands 3-5,
helix 2 and strand 3, and a few others, as seen in the second panel.
During the rotation of this loop values of RMSD and RG steadily grow.
The contacts inside loop 4 break temporarily to provide sufficient space to accommodate N and C terminal,
which are subsequently threaded through this loop, and which eventually leads to a slipknot conformation.
The colors of various elements of the structure in the bottom panel match
corresponding contact numbers in the top panel.
This folding process is shown in the attached movie, 
whose details are described in \emph{Appendix}.
} \label{backtracking}
\end{center}
\end{figure}

The fourth folding route is different from the ones mentioned above. 
Similarly to the previous cases, this route involves a creation of a knot close to N terminal and
a slipknot that involves terminal C.  At the late stages of folding however it also requires 
a huge rotation of the knotting loop (denoted by loop 4 in Fig. \ref{backtracking}). 
Both the knot and the slipknot are formed almost simultaneously during this surprising rotation of loop 4 by almost 360 degrees,
as shown by three steps in Fig. \ref{backtracking}. This one move makes the protein structure 
non-trivial. It has to be pointed out that some deviations of this route are also possible, 
but a critical role is always played by  rotation of loop 4. 
More details about the trajectories described above are given in the \emph{Appendix}.
We suggest that this mechanism, involving a rotation of the loop together with the steps from figure 2,   
should be typical in folding bigger and more complicated knotted structures.



\section{Discussion and conclusions}

The analysis of folding trajectories for topologically nontrivial 
proteins allows us to explore possible mechanisms for creation of knots and slipknots. 
Furthermore the results emphasize difficulties associated with these processes that 
commonly lead to kinetic traps. Below we summarize these two issues.

\subsection{Folding mechanism for nontrivial topologies}

Folding into topologically nontrivial configurations is associated to a series of complex events. 
Transitions between different stages in this process are nontrivial. In many cases they are not 
successful, which is associated with a presence of so called {\bf topological barriers}. When a protein 
does not manage to overcome such a barrier, it may be captured into an improper configuration -- a {\bf kinetic trap}.
Even successful events that overcome such a barrier may require a series of partial unfolding and refolding events,
which we call a {\bf backtracking}. Since backtracking is a stochastic process, it may lead to a broad 
distribution of time scales for such transitions.

The proteins YibK (1j85) and YbeA (1o6d, 1vho) have
the simplest type of knot, $3_1$. We determined that, in most cases, folding into this knotted structure 
goes through {\bf an intermediate configuration which contains a slipknot}, Fig. \ref{1o6d_path}.
A presence of this slipknot was conjectured in \cite{Taylor_2007,Yeates_2007},
and  our results confirm this prediction. A slipknot arises after threading a partially structured region,  
which we call a {\bf hook}, through a previously created {\bf loop}. 
The appearance of an intermediate slipknot in our simulations also suggests a plausible mechanism for 
the creation of deep knots (located far from both termini). 
We also expect that  more complicated types of knots (such as $4_1$ and $5_2$ \cite{Virnau})
are formed through a similar intermediate slipknot configuration.
Supporting this suggestion is the fact that the native state
of ubiquitin hydrolase UCH-L1 (PDB code 2etl) contains a sharp turn close
to its C terminus, which might be a remnant of an intermediate slipknot.

Folding into  knots and slipknots involves precise geometrical constrains. Knots and slipknots 
require threading a region of the protein through a loop. Protein chains, however, 
are not flexible strings since they are composed of helices and $\beta$-strands that typically do not freely bend. 
Since creation of a tight knot requires {\bf sharp turns},  
the protein needs to have properly located regions that allow for bending and motion. 
Locations of such regions cannot be accidental. For example, these sharp turns should occur around the loop region
to facilitate threading. As shown in Fig.~2, loop formation has to precede a sharp turn, 
otherwise correct loop formation would be impossible and 
would lead to kinetic trapping. Careful analysis of the protein structure already allows us 
to identify these bend regions before simulations are performed.

\subsection{Topological barriers and kinetic traps}

We suggest that the folding mechanism in Fig. 2 provides a route that reduces the
topological barriers in the free energy landscape of knotted proteins. These 
topological barriers have already been proposed earlier from an analysis of the structure (topology) of
various proteins \cite{norc}.
In our folding studies, these topological constraints due to  the presence of knots
severely restrict possible folding routes.
Topological constraints strongly restrict the folding landscape \cite{Onuchic_2000,Onuchic_2004}
and only allow one or a few similar trajectories leading to the native state.
An improper step along each such pathway may lead to a kinetic trap,
which corresponds to a local minimum in the folding landscape.
For shallow traps, the protein can resolve this kinetic problem with the aid
of backtracking.

Our numerical analysis has identified  various types of kinetic traps.
Recall that the geometry of a knot or a slipknot requires the formation of
sharp turns or flexible regions in appropriate places. These sharp turns often arise on prolines
or glicynes. Turns at amino acids 102, 114 and 125 in 1o6d in Fig. \ref{1o6d_path}
exemplify these regions. Turns are also formed at amino acids Pro and Gly (number 62 and 63) in the
native conformation. These are late turns formed much later than the others during the folding event.
During folding, however, these turns may be formed in an incorrect time order and/or may be formed at 
additional prolines or glicynes. In this situation the protein is unable to fold in the correct conformation. For example, 
when the turn at position 62 and 63 in 1o6d is formed too early, it precludes
the correct route towards folding.

The simple appearance of an intermediate configuration with a slipknot during the folding event does 
not yet guarantee that it will be followed by the formation of a knot. 
The hook, which is required to move into the  previously formed  loop,
may turn back and leave the loop, as shown in  the fourth step of Fig.
\ref{1o6d_path}. Successfully crossing of the loop by the terminus is needed to overcome the
topological barrier and therefore form the correct knot. Although we have successfully folded these proteins several 
times,
these multiple geometrical requirements are difficult to satisfy, and lead to kinetic traps on many occasions, 
particularly if the simulation temperature is too low. These challenges have inspired others to propose alternative 
mechanisms to
facilitate folding such as placing attractive long range non-native contacts \cite{shak} or 
chaperones to help push the hook through the loop. 
We submit that the use of non-native contacts prevents 
the crucial formation of an intermediate slipknot conformation.  
Our studies have the advantage of being based only on interactions associated
to native contacts. This allows to observe and to understand the role of an intermediate slipknot
in an unbiased way. Nonetheless, physically-motivated properly place non-native contacts, which do not 
interrupt the slipknot formation, may improve the success  rate of
folding events. This  point is worth further study.

To further understand the effect of the knot, we rebuilt protein 1o6d into a
very similar structure, but without knot topology \cite{Sulkowska_2008e}. 
As expected, under similar simulation conditions, the  number of successful
folding events increases substantially to about 83\%.
This shows that knot formation is the main limitation for folding, 
since in both cases the amino acid sequence is very similar, but 
the folding success rate is remarkably different. This provides direct 
evidence that the topological barrier arises as a consequence of knot formation.

Contrary to some previous experimental suggestions \cite{mallam_2008}
that knot formation is preceded by a formation of random knots, we do not
observe such a mechanism.
Indeed, the analysis of protein conformations during early stages of folding clearly shows
a noticeable number of randomly knotted structures.
Such behavior agrees with results of simulations in \cite{Virnau_2005} and well-known
experimental results that flexible polymers or strings \cite{Raymer} can easily become spontaneously knotted.
But in most cases random knots observed in folding simulations do not lead to deep but relaxed knots.
Notice that at room temperature knots on a protein chain
do not necessarily behave in the same way as in a polymer chain, which was
shown in the case of stretching simulations \cite{Sulkowska_2008}.
Additionally our results show that the knotting mechanism is not prevented even 
when additional tails are attached to the protein, which was also observed in experiment \cite{mallam_2008}. 
Without an intermediate slipknot it would be hard to explain the creation of knots in this case.
These results further support the mechanism described in this paper.

Finally, it would be more satisfying to achieve a higher success rate
of folding than just 1-2\% in our simulations. Such a small success rate may be a
consequence of several factors. Our folding routes may be too short.  Some
preliminary results indicate a higher degree of success for simulations five times longer.
Also the simplicity of the model that includes only native interactions and no geometrical details 
may cause limitations. Even with these limited coarse-grained models, however, we have already been able to 
provide strong insights about the possible mechanism governing folding in knotted proteins that can now
be checked by experiments and more detailed/expensive simulations.

\section{Materials}

\subsection{Proteins with knots and slipknots studied}

Knots observed in proteins are ''open'' knots, and so they differ from the mathematical definition of (closed) knots. 
Nonetheless, when both termini of the protein are
located far enough from its entangled core, they usually can be unambiguously joined by
an additional interval which transforms them to a closed loop. If such a loop is not
homeomorphic to a circle then the native protein is regarded as representing a nontrivial (open)
knot. The families of proteins with a trefoil knot $3_1$ studied here are
YibK (PDB code 1j85) and YbeA (including structures with PDB code 1o6d, 1vho).
The slipknots are topological configurations more subtle than knots. They exist when
a piece of a protein chain gets in and out of a loop formed by some other part of the chain.
This means that cutting several residues form one end of the protein would lead to a configuration
with an (open) knot, which is absent in the native state. One protein with a slipknot which we study is
a highly conserved protein from crenarchaeal viruses AFV3-109 \cite{Keller} (PBD code 2j6b).
It consists of N=109 amino acids, belongs to $\alpha/\beta$ class and is built of
five $\beta$ strands, which form the sheet surrounded by a loop from one side and
the helices on the other side, Fig. \ref{zwijanie-rys-slip}. This protein forms a dimer with the shape
of a cradle 
\cite{Keller}.
2j6b contains a slipknot, such that removing 9 amino acids from C terminal side
leads to a knotted configuration. The smooth topological representations of
2j6b is shown in Fig. \ref{zwijanie-rys-slip}.
The function of 2j6b is still unknown, however it was suggested that it
could interact with nucleic acids \cite{Keller}.
The second protein with a slipknot which we study is a thymidine kinase \cite{Gardberg} (PDB code 1p6x)
which belongs to $\alpha/\beta$ class, consists of $\beta$ sheet of 5 beta strands which
are surrounded by 6 helices,  and a bundle of 5 periferial $\alpha$ helices, Fig. \ref{zwijanie-rys-slip}.
It contains a slipknot, such that considering only amino acids 26-140 one finds
a knot $3_1$ made of 3 $\beta$ sheets and 3 helices. The slipknot arises when the bundle of helices
and 2 $\beta$ sheets come back to the structure by the loop between 121 and 132 amino acids.
As was described in \cite{Yeates_2007} the slipknot is very deep on C terminal side, however
it is shallow on the N terminal side, with only 10 residues extending
out of the knot core. This difference suggests that the knot is most likely formed
be the N terminal segment of the protein passing through a loop in the protein arising
during the later stages of the protein folding. Because the N terminal is shorter, the slipknot
in thymidine kinase is considerably more shallow than the one in alkaline phosphatase;
this knot is also much tighter. We use thymidine kinase to analyze its folding ability
and mechanical properties. The analysis of protein structure is described in \emph{Appendix}.\\
\subsection{Reidemeister moves} Three Reidemeister moves shown in Fig. \ref{pathway-all} describe basic geometric
transformations which do not change a type of a knot.
They are very useful in a simplified description of proteins with nontrivial topology.\\
\subsection{Coarse-grained model} We use the coarse-grained molecular dynamics modeling
described in detail in refs. \cite{Cieplak_2003,Cieplak_2004,Cieplak_2004a}.
The native contacts  between
the C$^{\alpha}$ atoms in amino acids $i$ and $j$ separated by the distance $r_{ij}$
are described by the Lennard-Jones potential
$V_{LJ} = 4\epsilon \left[ \left( \sigma_{ij}/r_{ij}
\right)^{12}-\left(\sigma_{ij}/r_{ij}\right)^6\right]$.  The length
parameter $\sigma _{ij}$ is determined pair-by-pair so that the minimum
in the potential corresponds to the native distance.
The energy parameter $\epsilon$ is taken to be uniform. As discussed in
ref. \cite{Sulkowska_2008d}, other choices for the energy scale and the form
of the potential are either comparable or worse when tested against experimental data on stretching.
Implicit solvent features come through the velocity dependent damping 
and Langevin thermal fluctuation in the force.
We consider the over-damped situation which makes the characteristic
time scale, $\tau$, to be controlled by diffusion and not by ballistic-motion,
making it of order of a ns \cite{Szymczak_2007}.
Thermodynamic stability of a protein can be characterized by providing the folding temperature 
$T_{f}$ at which half of the native bonds are established on average in an equilibrium run 
(based on at least ten long trajectories that start in the native state).
The analysis of the knot-related characteristics is made along the lines described in ref. \cite{Sulkowska_2008}. \\
\subsection{KMT algorithm} We determine the sequential extension of a knot, i.e.
the minimal segment of amino acids that can be identified as a knot,
by using KMT algorithm \cite{km1}. It involves removing the
C$^{\alpha}$ atoms one by one, as long as the backbone does not intersect a
triangle set by the atom under consideration and its two nearest sequential
neighbors.


\section{Appendix}

In this Appendix we present more details concerning the geometry and properties of knotted
proteins (1j85, 1o6d and 1vho) and proteins with slipknots (2j6b and 1p6x). 
We present the Reidemeister moves necessary for a description of knot creation and describe attached video clips.
Furthermore, we provide quantitative details of the fraction of native contacts, as well as the "optimal" energetic 
map which we found. Finally we describe in detail four folding trajectories of 1p6x.

\appendix

\section{Proteins with knots and slipknots }

\subsection{Knotted proteins}

\begin{figure}[htb]
\begin{center}
\includegraphics[width=0.9\textwidth]{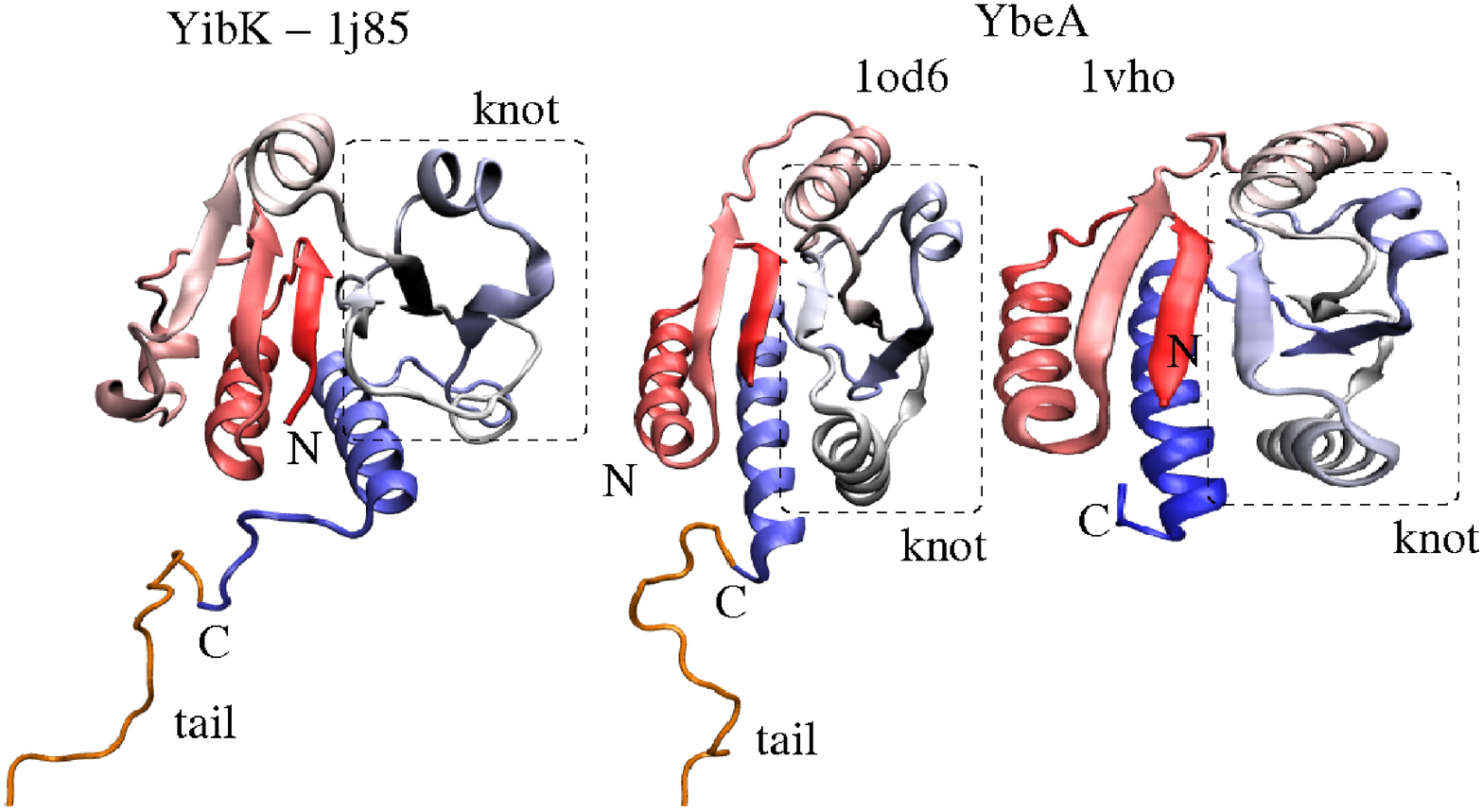}
\caption{Left: 1j85. Middle: 1o6d. Right: 1vho. For each protein a region where the knot is localized
is denoted by a dashed rectangle.}
\label{bialka}
\end{center}
\end{figure}

Monomeric subunits of haemphiluse influenzae YibK (PDB code 1j85) and escherichia coli YbeA (PDB codes 1o6d and 1vho) 
are shown in Fig. \ref{bialka}, together with schematic diagrams representing their topology. 
Another member of the methylotransferase YibK, the protein sr145 
from Bacillus subtilis (1PDB code 1to0), is not suitable for our analysis 
as its chain is broken in many places.

Metylotransferase (Mtases) from Hamophilus influenazae YibK is
described in detail in \cite{mallam_2007a,mallam_2007b}.

Homodimeric $\alpha/\beta$-knot metylotransferase (Mtases) from escherichia
coli YbeA (1o6d),  consists of 5 $\alpha$ helices which surround
the $\beta$ sheet built of 6 $\beta$ strands.
The secondary structure elements are referred to as follow:
1 -- $\beta$ strand (amino acids 2-8);  2 -- $\alpha$ helix (12-27);
3 -- $\beta$ strand (31-37); 4 -- $\alpha$ helix (44-58);
5 -- $\beta$ strand (65-69); 6 -- $\beta$ strand (74-75);
7 -- $\alpha$ helix (77-92); 8 -- $\beta$ strand (95-99);
9 -- $\alpha$ helix (106-113); 10 -- $\beta$ strand (115-118); 
11 -- $\alpha$ helix (125-145).
The structure is knotted between $\beta$ strands 5 and 10.

The second protein from the YbeA family (1vho) has almost the same structure as 1o6d, 
however it is longer by 10 amino acids. 
It consists of 5 $\alpha$ helices which surround
the $\beta$ sheet built of 6 $\beta$ strands.

\subsection{Proteins with slipknots}

We analyse two proteins with slipknots:  thymidine kinase \cite{Gardberg} (PDB code 1p6x)
and highly conserved protein from crenarchaeal viruses AFV3-109 \cite{Keller} (PBD code 2j6b). 

\begin{figure}[htb]
\begin{center}
\includegraphics[width=0.4\textwidth]{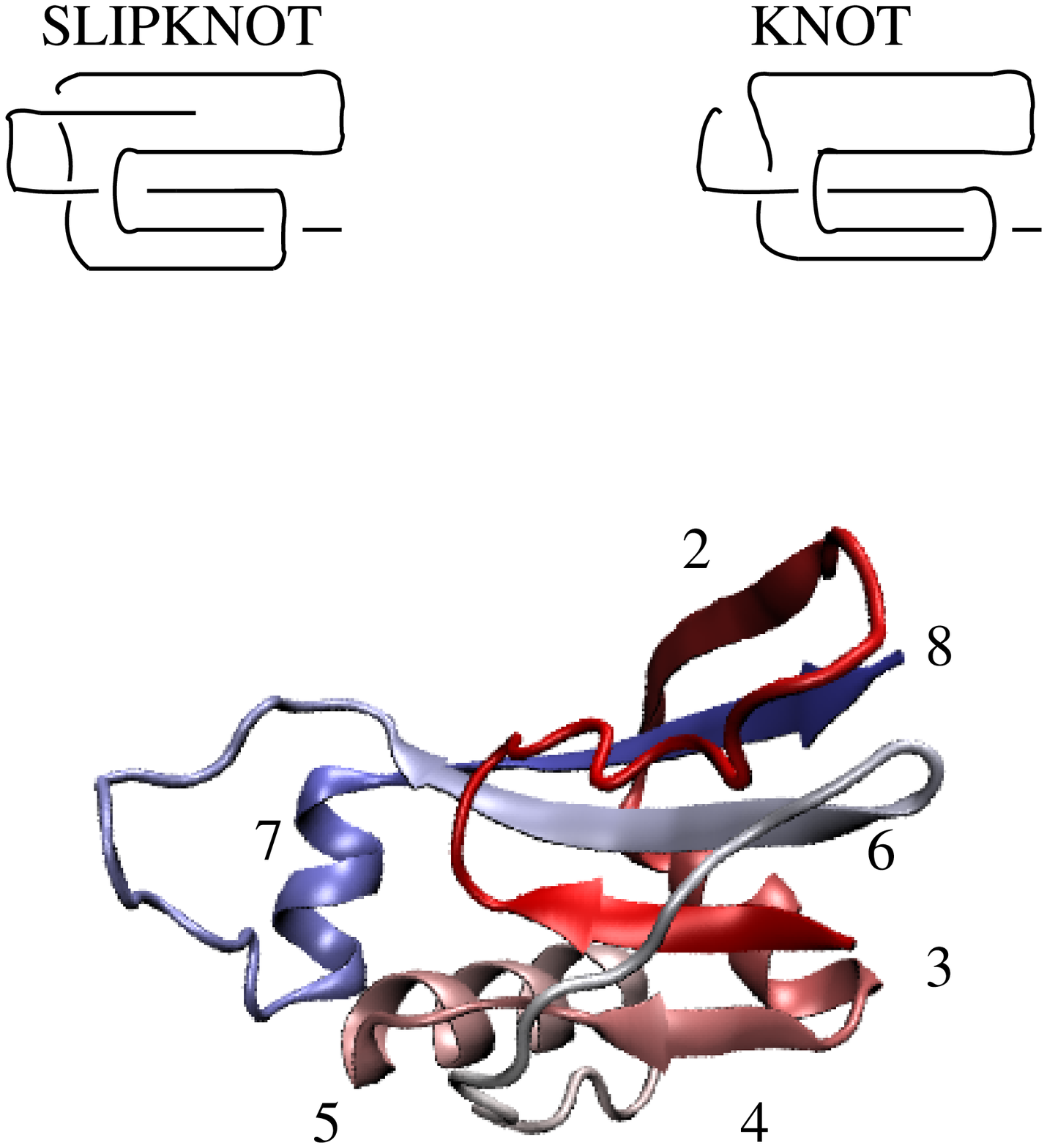} \hspace{1cm}
\includegraphics[width=0.4\textwidth]{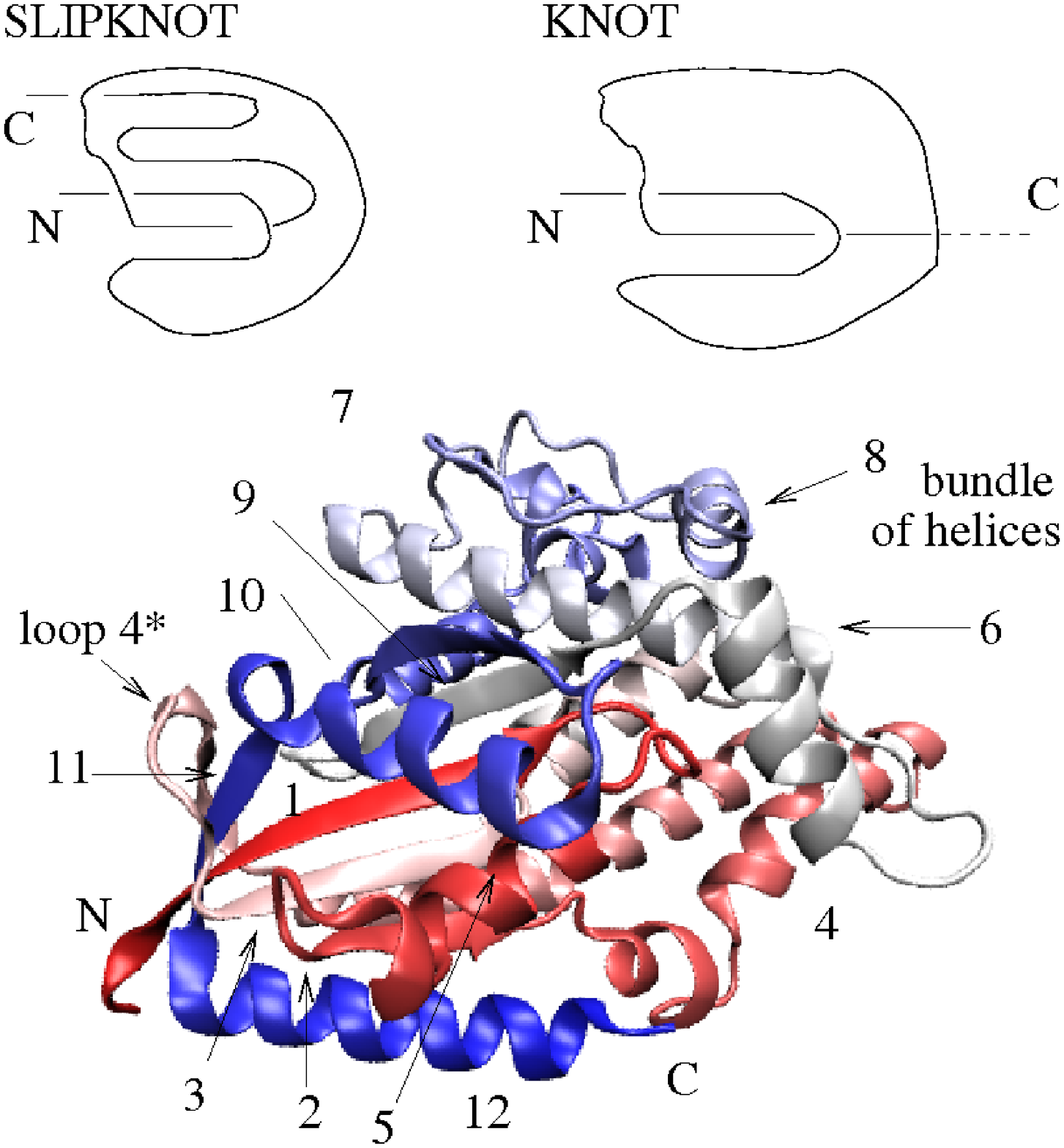}
\includegraphics[width=0.99\textwidth]{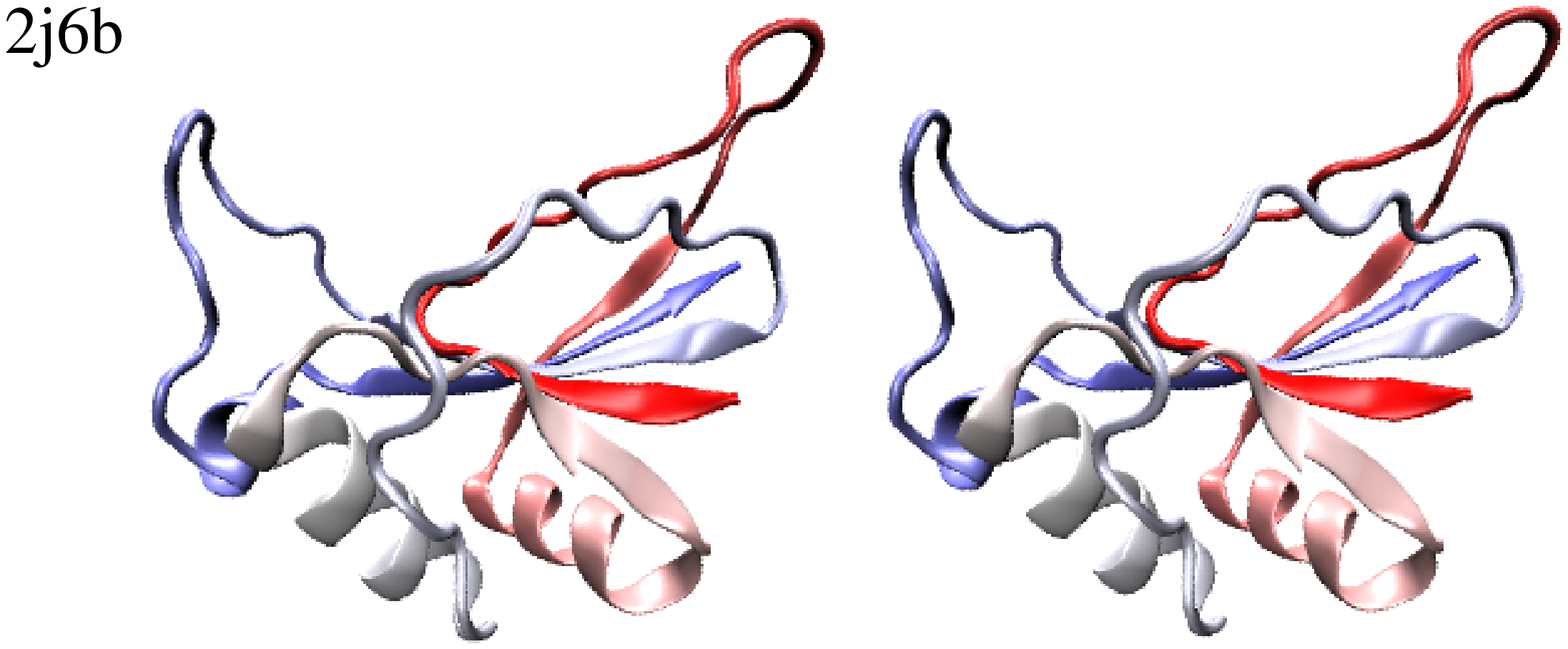}
\includegraphics[width=1.4\textwidth]{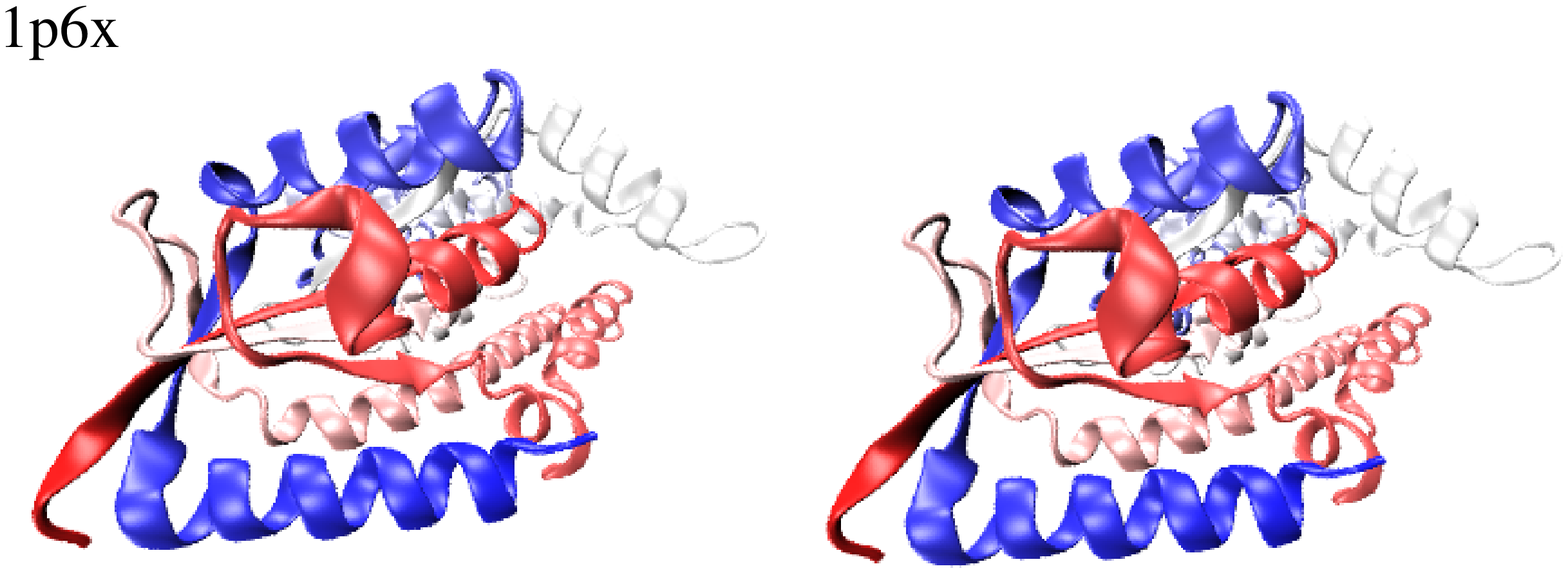}
\caption{Bottom: cartoon structures of proteins 2j6b (left) and 1p6x(right) . Top: 
the corresponding schematic structures of a slipknot in these proteins, as well as
the knotted conformation $3_1$ obtained after cutting, respectively, 
20 amino acids from the N terminal (for 2j6b), or a few amino acids from C terminal (for 1p6x). 
The lowest figures stereoview of proteins 2j6b and 1p6x.
} \label{fig-2j6b-1p6x}
\end{center}
\end{figure}

Carton representation of the protein 2j6b is shown in Fig. \ref{fig-2j6b-1p6x} on the left, 
together with a schematic structure of a slipknot, 
as well as a knot obtained after cutting 9 amino acids from the terminal C. 
2j6b consists of N=109 amino acids and belongs to $\alpha/\beta$ class. 
It is built of five $\beta$ strands   
which form the sheet surrounded by a loop from one side and the helix 3-5 on the other side. 
The secondary structure notation which we use is as follow: 
1 -- $\beta$ strand (amino acid 2-5);  2 -- $\beta$ strand (19-25);
3 -- $\alpha$ helix (26-37); 4 -- $\beta$ strand (39-41);
5 -- $\alpha$ helix (45-57); 6 -- $\beta$ strand (74-80);
7 -- $\alpha$ helix (92-100); 8 -- $\beta$ strand (101-109).
The function of AFV3-109 is not known, however it was suggested that it 
could interact with nucleic acids \cite{Keller}.

A thymidine kinase (PDB code 1p6x) belongs to $\alpha/\beta$ class protein and 
is shown in Fig. \ref{fig-2j6b-1p6x} on the right.
It can be divided into 11 parts denoted by numbers from 1 to 11 in the main text, which are specified as follows: 
1 -- N terminal with the $\beta$ strand (amino acid 20-34); 2 -- $\alpha$ helix with turn 
(35-45); 3 --$\beta$ hairpin and turn (55-60); 4 -- big long loop created of 3 helices form one side and 
closed from the other side by loop 4* (120-132) all toghter (61-132); 5 -- $\beta$ strand (133-139);
6 -- hairpin built of two helices (141-176); 7 -- $\beta$ strand (177-184); 8 -- bundle of five helices 
(185-299); 9 -- $\beta$ strand (300-306); 10 -- helix and $\beta$ hairpin (306-323);
11 -- the chain crossing the loop 4* (324-333); 12 -- helix on the second side of the loop 4 (334-352).
 The strands 1, 3 and 5 create a characteristic $\beta$-sheet.
Using this notation for the elements of the secondary structure we identified the following contacts: 
1-2, 1-4, 1-4*, 1-5, 1-7, 1-10, 1-11b, 2-3, 2-10, 3-5, 3-10, 3-12, 4, 4-5, 4-6, 4-8, 4-10, 
4-11, 4-12, 5-12, 6, 6-8, 7-9, 7-10, 8, 9-10.
This list is used to analyze the folding routes and the backtracking between various groups of contacts. 
Our analysis is mostly concentrated on the loop 4* through which the $\beta$ strand 1,  
$\beta$ strand 10, and helix 11 have to pass. 

\section{Reidemeister moves}

\begin{figure}[htb]
\begin{center}
\includegraphics[width=0.95\textwidth]{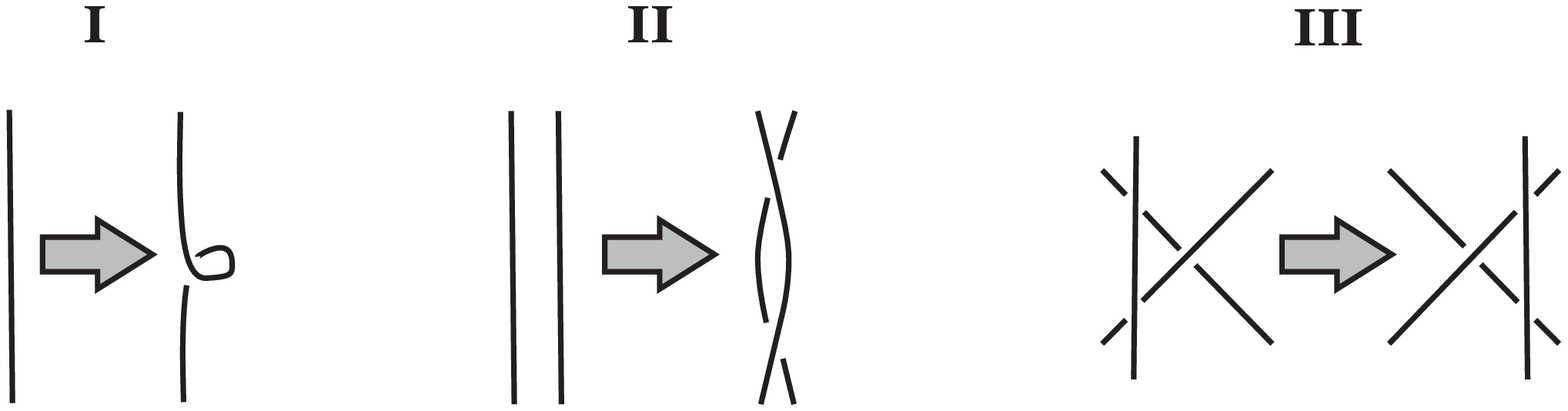}
\caption{Reidemeister moves I, II and III. 
} \label{reide}
\end{center}
\end{figure}

Three Reidemeister moves I, II and III are shown in figure \ref{reide}.
They describe, after a projection on a plane, basic geometric transformations 
which do not change a topology of a knot. In each move the relative
location of the ends of all the strings involved is unchanged. 
The move I is a transformation of a straight piece of a string
into a loop. The move II corresponds to transforming two parallel strings 
into a configuration with two crossings. The move III involves three strings,
and corresponds to a shift of one of them over the crossing made by the other two.
These Reidemeister moves are very useful in a simplified description of proteins with nontrivial topology. 

\section{Video clips of folding trajectories for knots and slipknots}

We enclose a video presentations of the mechanism of knot formation,
generated using our implementation of the Go-like model.
The first animation presents protein 1j85, and the second
the protein 1o6d. The process of the knot formation in these
animations corresponds to Fig. 1 in the main text. 
The knotted regions of both proteins are shown by green color. 
Furthermore, we also enclose a video presentations of the mechanism of slipknot formation 
in our model, respectively for 2j6b and then 1p6x.

\section{The intermediate slipknot and the fraction of native contacts for 1o6d}

\begin{figure}[htb]
\begin{center}
\includegraphics[width=0.35\textwidth]{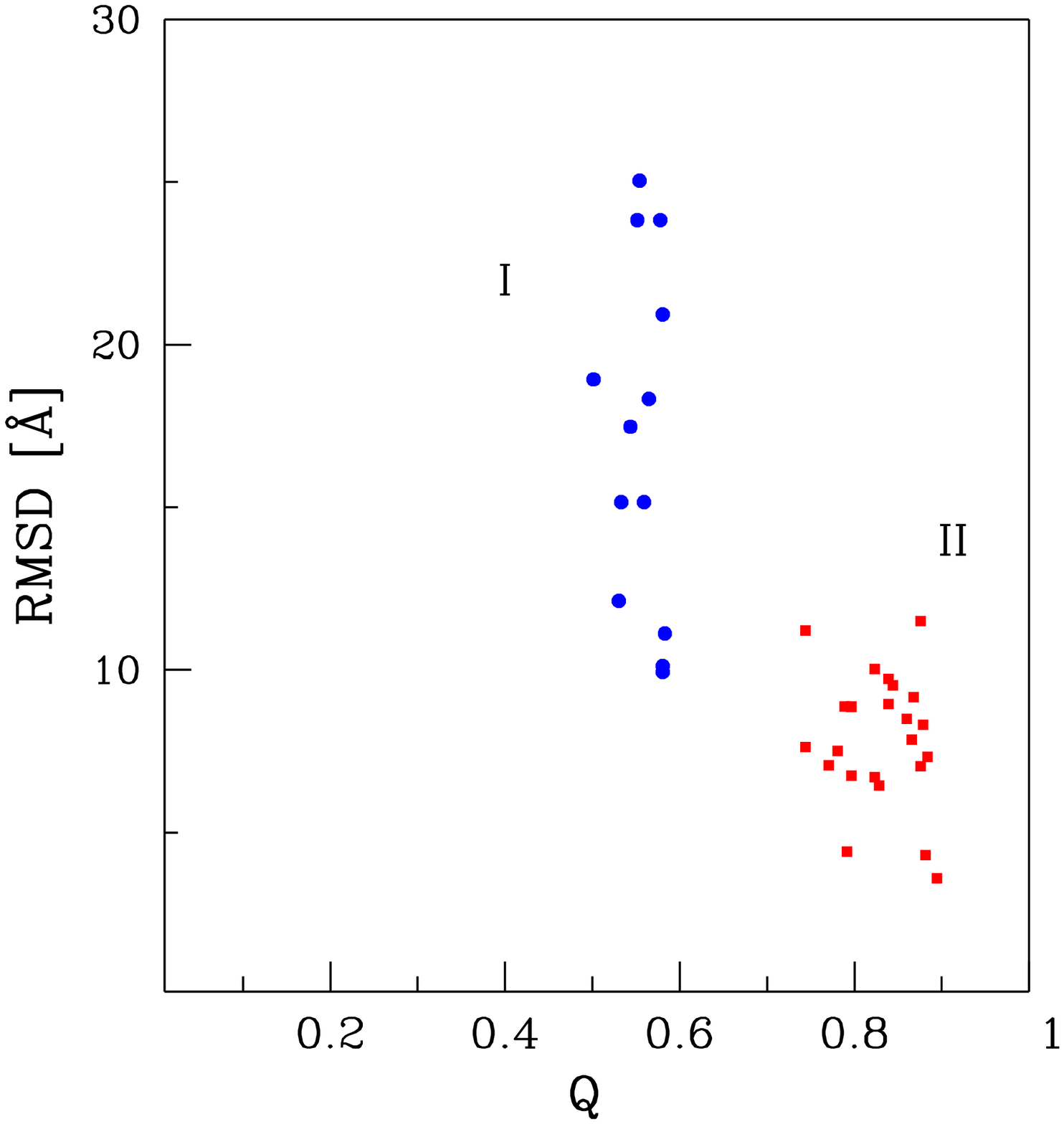}
\caption{Distribution of native contacts in protein 1o6d
at the moment when slipknot configuration appears for the first time during folding.
This moment corresponds to the fourth step in Fig. 2 in the main text.
The results are similar both for the model with the optimal native like energy scale,
and the basic model with uniform energy. The symbol I corresponds to folding from
the terminal C, while II to folding from terminal N.}
\label{rozklad_q}
\end{center}
\end{figure}

In Fig. \ref{rozklad_q} the fraction of native contacts $Q$ is shown at the moment when
the slipknot is being created, i.e. when its hook-part starts to be threaded through the loop 66-96.
In the main trajectory the structure elements located closer to the N terminal
of the protein fold in last stages of folding process. These trajectories are represented
by red dots and in this case $Q \simeq 0.8$.
In the second type of trajectories, represented by blue dots,
the N terminal folds in initial stages of the folding, and $Q \simeq 0.6$.
One additional point in the figure represents a very rare knot formation,
corresponding to threading the terminus C through the loop 66-96 without a slipknot intermediate.
The data shown in this figure corresponds to the optimal model considered in the main text and
denoted in bold in the table below. However, there is no qualitative difference in the values of $Q$ 
between this model and the results obtained in the basic model (with the uniform energetic map).

Apart from the protein 1o6d we also checked the folding ability of another member of YbeA family, i.e. protein with PDB 
code 1vho.
Similarly like 1o6d, this new protein can fold to the native state in the model only based on the native contacts. 
The folding route includes the slipknot transition state, similarly as for 1o6d and 1j85. 
However for this protein we also observed one trajectory where the C terminal crosses the loop 
not in a hairpin-like configuration.  
The probability of creation of a knot for this protein in a simple model is around 0.1\%.

\section{New energetic map for folding knotted protein 1o6d}

From the analysis of the behavior (in particular involving the backtracking)
of various elements of the knot structure in 1o6d during folding, 
shown in Fig. 3 in the main text, we constructed a series of models with new energetic maps. 
These maps are based only on native contacts and differ in strengths of interactions
inside the knotted region of a protein. The energetic maps which we considered are shown in table 1,
where these groups of contact are listed whose strength was changed.
We found a few energetic maps for which the 
folding ability is highest -- they are shown in rows 15, 17 and 20 in the table 1. 
For these models we increased twice the sampling set, and under such refined statistics
we found that folding ability was the best 
for the model from the row 15, which we call "optimal".
It is also possible to construct energetic maps 
which prevent knot formation completely, for example by increasing the strength of 
interaction between contacts for which backtracking is observed (such as $\beta$ strands 5, 6 and 9).

\section{Folding trajectories for slipknots in 1p6x}

\begin{figure}[htb]
\begin{center}
\includegraphics[width=0.59\textwidth]{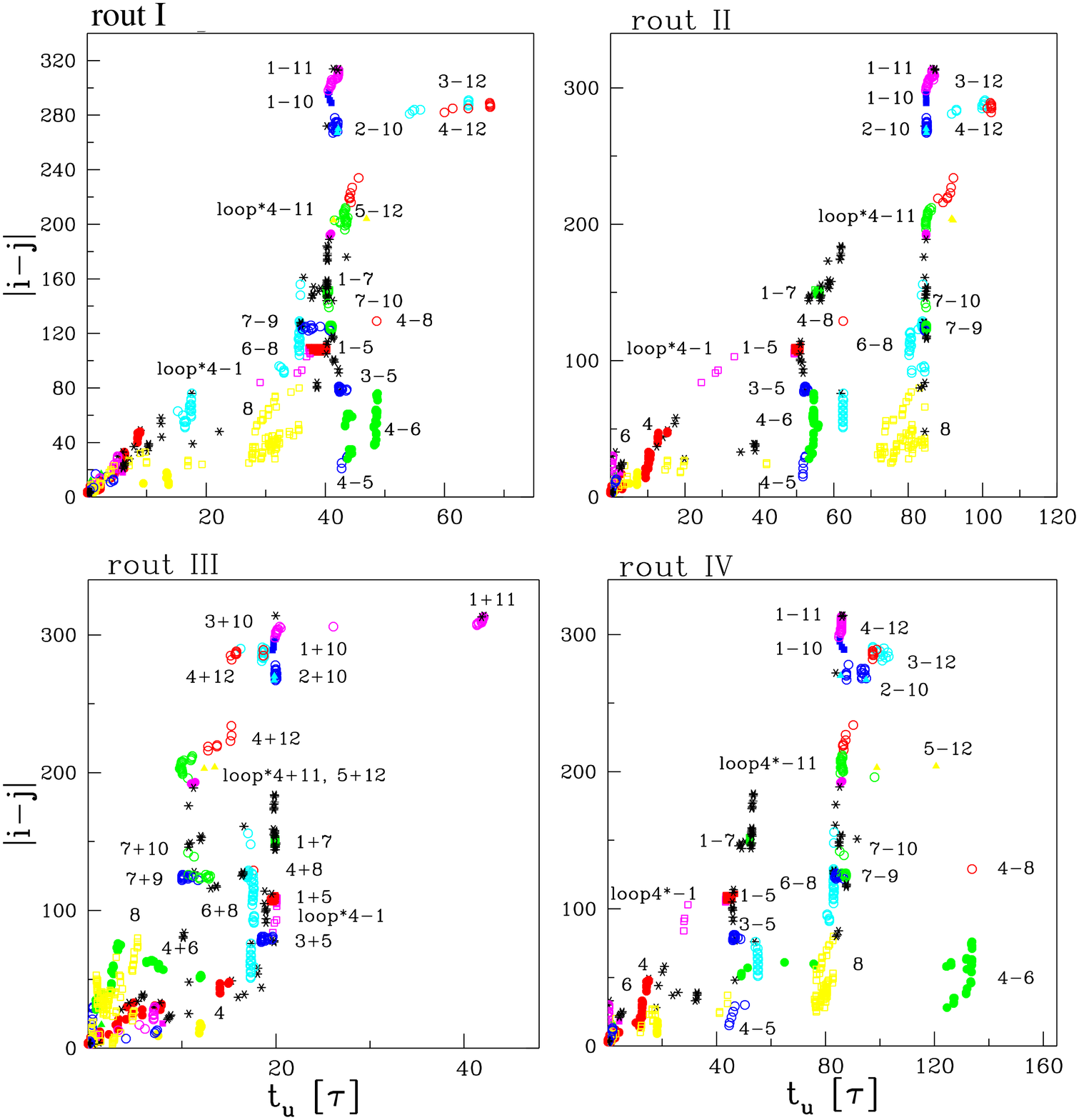}
\caption{The folding routes for the 1p6x protein as described
by the first time of forming a contact corresponding to the sequence
of length $|j-i|$, respectively for the first, second, third and fourth route.
} \label{pathway-all}
\end{center}
\end{figure}

We present now in more detail four main folding routes shown 
in Fig. \ref{pathway-all} which are briefly discussed in the main text.

The first route is represented by 5 trajectories which we found.
Formation of the slipknot is one of the last steps during folding, and it is possible due to backtracking
between $\beta$ strands 1-5, 3-5, and contacts between 3-11 and 4-5.
The characteristic property of these trajectories is 
threading of the terminals N and C through the loop $4^*$,
which takes place after or almost in the same time
when the entire loop 4 establishes contacts with the segment 8. 
This event, as well as a correlation between threading almost in the same time the elements
1 and 10, makes this trajectory distinct from all others which we observed.
From folding scenario for this route we deduce, that almost all native contacts are created before
threading appears, apart from the contacts between N terminal
and $\beta$ hairpin of C terminal with loop $4^*$.
At this state the contacts between 1-4, 1-5, 4-5 and 4-6 have to be broken
to open the loop 4 and allow to thread N and C terminals through it.
Now a hairpin is created close to the terminal N, which subsequently moves across the loop 4.
By temporary breaking and creating the contacts between terminal N and hairpin 10,
the entire C terminal is threaded across the loop 4 also in the hairpin conformation.

The second folding route is represented by 2 trajectories.
In this case the chain strongly folds from N terminal by a few different nuclei.
This causes creation of a $\beta$ sheet from $\beta$ strands 1, 3, 5,
entire loop 4 and its contacts with 6 and 8. In such case (no other native structure of protein is yet established)
a small backtracking between elements 1-4 and 1-5 allows easily for crossing terminal N in the bend shape
through the loop $4^*$ which creates a knot. After that all other elements of the protein fold.
The final event is threading the terminal C across the loop $4$ as in the first folding route described above.

The third folding route is most distinct form others.
A part of the loop 4, loop 6 (made of two helices) and bundle 8 fold in the same
time when contacts between 9 and 10 are established, however all nuclei still exist separately.
In the next step big nuclei start to rearrange, which results in a creation of a slipknot 
with the segments 9, 10 and 11 located inside the loop 4.
In contrast to all other trajectories, after this step the N terminal starts to fold
and makes half of the $\beta$ sheet clamping the C terminal in the loop 4.
Now only the end of the terminal N has to be threaded by the loop $4^*$ (as in the route 1),
while the terminal C has to go out of the clamp (as in the route 2).
These moves are possible due to backtracking between the segments  1-4, 1-5, 4-5, 1-10 and 3-10.

The fourth trajectory is schematically described in the main text, however now we include more details.
We remind that in this route almost entire native structure, apart from contacts between the loop 4 and elements
6 and 8, is created before creation of a slipknot.
This conformation corresponds to almost entirely created $\beta$ sheet (involving the elements 1, 3 and 5)
with strands 1 and 10 located close to their native positions (as in the knot or slipknot configuration).
The final stages of knotting are accompanied by a backtracking, as
shown in Fig. 6 in the main text.
We discuss it starting from the moment when the loop 4 is still above the sheet 1-3-5.
From this moment the contacts between $\beta$ strands 3-5,
helix 2 and strand 3, and a few others, perform amazing backtracking
which leads to the rotation of the loop 4 by almost 360 degrees. During this rotation
the values of RMSD and RG steadily grow.
The contacts inside loop 4 break temporarily in order to provide enough space to accomodate
both terminals N and C, which are subsequently threaded through the loop.

Some deviation of the second folding route is represented by two other trajectories,
for which the critical role is also played  by closing the loop 4. 
In this case the $\beta$ sheet 1-3-5 is created after establishing entire remaining native structure
and before closing the loop 4 (as also described above). This allows for reopening the
contacts in $\beta$ sheet and rotation of loop 4 to make the knot, despite
somewhat awkward position of the terminal C. In consequence closing of the loop 4
clamps the terminal C between this loop and the $\beta$ sheet for a long time.

\bigskip

\bigskip

\begin{center}
\begin{Large}
{\bf Acknowledgments}
\end{Large}
\end{center}

\medskip

We appreciate discussions with M. Cieplak, P. A. Jennings, P. Szymczak, and P. Wolynes.
We thank D. Gront for help with reconstructing the proteins.
This work was supported by the Center for Theoretical Biological
Physics sponsored by the NSF (Grant PHY-0822283) with additional
support from NSF- MCB-0543906.
P. Su{\l}kowski highly appreciates the support of the Humboldt Fellowship
and great hospitality of the University of California San Diego.

\begin{table*}
\caption{
}
\begin{center}
\small{
\begin{tabular}{c|c|c|c|c|c|c}
\hline
Order & region*$\epsilon$  & region*$\epsilon$     & region*$\epsilon$ & region*$\epsilon$ &   success rate    \\ 
      &                    &            &         &  &                           \\
\hline
1. &  5-8*0.5             &            &    			&    		&    3   \\
2. &  5-8*2             &            &    			&    		&    9   \\
3. &  5-8*4             &            &    			&    		&   10   \\
4. &  5-8*6             &            &    			&    		&    9   \\
5. &  5-8*10            &            &    			&    		&    1   \\
   &                    &            &    			&    		&        \\
6. &  5-8*2             &  6-10*2    &    			&    		&      15  \\
7. &  5-8*4             &  6-10*2    &    			&    		&       9  \\
8. &  5-8*10            &  6-10*2    &    			&   	        &      12  \\
   &                    &            &    			&    		&          \\
9. &  5-8*2             &  6-10*4    &    			&    		&      14  \\
10.  &  5-8*2            &  6-10*6    &    			&    		&       8  \\
11.  &  5-8*2.5          &  6-10*2.5  &    			&    		&       6  \\
    &                    &            &    			&    		&           \\
12. & 5-8*2              &            & (7-10,8-10,9-10)*0.5   &    		&      2   \\
13. & 5-8*2              &            & (7-10,8-10,9-10)*1.5   &    		&      2     \\
14. & 5-8*2              &  6-10*2    & (7-10,8-10,9-10)*0.5   &    		&     25     \\
\bf{15.} & 5-8*2          &  6-10*2    & (7-10,8-10,9-10)*0.5   & without 74-115  & \bf{34}   \\
17.  & 5-8*2.5            &  6-10*1.5  & (7-10,8-10,9-10)*0.5   & without 74-115  & 25   \\
18.  & 5-8*2              &  6-10*3    & (7-10,8-10,9-10)*0.5   & without 74-115  & 27   \\
19.  & 5-8*1              &  6-10*3    & (7-10,8-10,9-10)*0.5   & without 74-115  & 15   \\
     &                    &            &                        &                &       \\
20.  & 5-8*2              &  6-10*2    & (7-10,8-10,9-10)*0.5   & turns*2        & 34  \\
       &                    &            &                        &                &       \\
21.    & 5-8*2              &            &                        & 5-9*2             &   3    \\
22.    & 5-8*2              &            &                        & 6-9*2            &   2    \\
\end{tabular}
}
\end{center}
\label{tab-F-d}
\end{table*}

\end{document}